\documentclass[conference]{IEEEtran}
\IEEEoverridecommandlockouts
\usepackage{hyperref}
\usepackage{cite}
\usepackage{amsmath,amssymb,amsfonts}
\usepackage{algorithmic}
\usepackage{graphicx}
\usepackage{textcomp}
\usepackage{xcolor}
\def\BibTeX{{\rm B\kern-.05em{\sc i\kern-.025em b}\kern-.08em
    T\kern-.1667em\lower.7ex\hbox{E}\kern-.125emX}}
    
\begin{document}

\title{Heterogeneous Federated CubeSat System: problems, constraints and capabilities\\
\thanks{Funding CAPES process number 88882.444451/2019-01}
}

\author{\IEEEauthorblockN{%
Carlos Leandro Gomes Batista\IEEEauthorrefmark{1}, 
Fatima Mattiello-Francisco\IEEEauthorrefmark{1}, 
Andras Pataricza\IEEEauthorrefmark{2}}%
\IEEEauthorblockA{%
\IEEEauthorrefmark{1}%
Brazilian National Institute for Space Research, \\
Space Systems Engineering,\\
São José dos Campos, Brazil \\
Email: \{carlos.batista, fatima.mattiello\}@inpe.br%
\IEEEauthorblockA{%
\IEEEauthorrefmark{2}%
Budapest University of Technology and Economics \\
Department of Measurement and Information Systems\\
Budapest, Hungary \\
Email: pataricza.andras@vik.bme.hu%
}
}
}

\maketitle
\begin{abstract}
Different arguments were being presented in the last decade about CubeSats and their applications.
Some of them address wireless communication (5G and 6G technologies) trying to achieve better characteristics as coverage and connectivity.

Some arrived with terms as IoST (Internet of Space Things), Internet of Satellites (IoSat), DSS (Distributed Space Systems), and FSS (Federated Satellite Systems).

All of them aim to use Small/NanoSatellites as constellations/swarms is to provide specific services, share unused resources, and evolve the concept of satellites-as-a-service (SaS).

This paper aims to emophasize performance attributes of such cyber-physical systems, model their inherent operational constraints and at the very end, evaluate the quality of service in terms of figures of merit for the entering/leaving of new heterogeneous constituent systems, a.k.a satellites, to the constellation.
This "whitepaper"-styled work focuses on presenting  the definitions of this heterogeneous constellation problem, aims at its main capabilities and constraints, and proposes modeling approaches for this system representation and evaluation.

\end{abstract}


\section{Introduction}
As the space became more and more accessible, new ways of thinking about the space services have also become more feasible.
Issues, like flight formation, constellations, and swarms of satellites were always desirable for the leading space agencies and enterprises.
GNSS \cite{hofmann2007gnss}, Sentinel \cite{attema2010sentinel}, and Iridium \cite{maine1995overview} are examples of well-established satellites constellation systems, using the coordination between specific purpose developed spacecrafts for global positioning, earth observation and IoT communication, respectively.

The rapid electronics advances, increase of processing capabilities, and low power consumption changed this discussion completely.
Since the introduction of the CubeSat standard \cite{cubesat2001puig}, many works aim the exploration and development of new capabilities for these small satellites \cite{book2016cubesat}.
Additionally, the decreased size of these platforms also reduced the launch costs \cite{Kulu2021, fox}.
Now, it is possible to, literally, launch hundreds of less than $5 kg$ satellites, each one with entirely different payloads, characteristics, owners, and purposes \cite{reuters_2017}.
All these points lead us towards a new era of re-thinking the idea of satellite constellation systems and their applications.

This work aims to evaluate the primary constraints, problems, and capabilities of such new ideas of using CubeSats in the context of decentralized ownership of heterogeneous satellites cooperating to achieve a common goal, what we can call a Federated CubeSat System.

\section{Concepts}

Distributed Satellite Systems, DSS, are defined as space systems that allocate functionality through multiple constituent systems to achieve a common goal \cite{Selva2012cubesatearth}.
These constituent systems are generally different spacecrafts with \cite{Cappelletti2021cubesathb}: 
(a) same capabilities (constellations and swarms) or; 
(b) fragmented functionalities, each satellite performs a different activity to achieve the main goal or;
(c) decentralized ownership, a federated satellite system (FSS), where different organizations contribute with new satellites and infrastructure.

More specific about the FSS, this kind of system-of-systems establishes the active sharing of unused resources of the multiple constituent systems offered for exploitation in different ways.
Shared resources from hosted payloads can service time, processing power, and data rate. Different integration concepts include Internet of Satellites, IoSat, and Internet of Space Things, IoST. They focus on communication and connectivity, relying on the involvement of the spacecrafts on what is called Inter-Satellite Networks, ISN, and Inter-satellite Links, ISL \cite{Golkar2013fss}.

Key enabling technologies \cite{Cappelletti2021cubesathb}, like:
(i) dynamic resource allocation and balancing;
(ii) power-efficient software-defined radios;
(iii) satellite negotiator;
(iv) software-defined satellite;
(v) virtual space missions,
should also be taken into consideration to make these kinds of systems possible.

Once the FSS has implemented the satellite services, its generated payload data can become commodities from the user's point of view, the satellite-as-a-service (SaS).
Here, we use the concept of FSS not only for spacecrafts, but we intend to expand the idea to the whole space system (space, ground, and user segments).

\section{Problems, Requirements \& Constraints}

The major problem in dealing with FSS is its lack of homogeneity \cite{Cappelletti2021cubesathb}.
Managing heterogeneity requires some common rules. 
Starting with a Systems Engineering approach, we must define such a system's primary needs, goals, objectives, and constraints.
The goal is to develop measures that will drive each constituent system's impact, with their particular capabilities.

\begin{figure}[h!]
    \centering
    \includegraphics[width=0.45\textwidth]{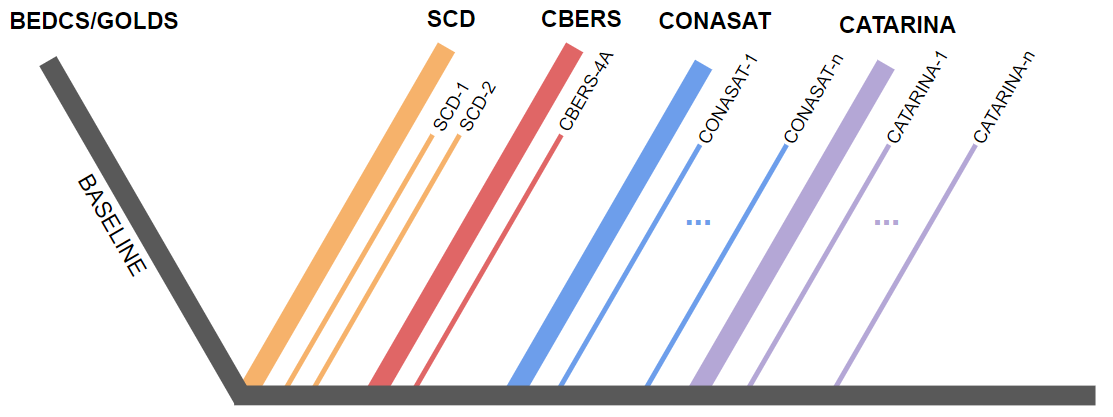}
    \caption{Baseline and possible composition view of the BEDCS/GOLDS SoS and its constituent systems.}
    \label{fig:baseline}
\end{figure}

Let us use as an example the Brazilian Environmental Data Collection System, BEDCS, \cite{yamaguti1994collection} and its exploitation as GOLDS -- Global Open coLlecting Data System \cite{vidal2021global}.
The BEDCS and GODLS serve identical overall needs, goals, and objectives, get the data from the Data Collection Platforms (DCP) spread around a specific territory, and send it to the Ground Stations. However, they differ in constraints due to the capabilities of their constituent systems (Figure~\ref{fig:baseline}).

For the BEDCS, we have the SCD and CBERS family satellites.
These satellites differ in form, design, mission, and primary objective.
The CBERS main goal is Earth Observation, and the Data Collection is secondary.
Nevertheless, they share the Ground Stations and the hosted payload, which requires simultaneous access/contact with DCP and Ground Station.

The GOLDS, on the other hand, intends to use a new generation of CubeSats from CONASAT and CATARINA constellations beside the SCD and CBERS satellites.
Each constellation has its particular objectives, but they will share a common goal for the GOLDS through a new hosted payload, the Environmental Data Collector (EDC), that enables the GOLDS to be global and overcome the simultaneous visibility constraint from BEDCS.
Turning the data collection global, we start to deal with new constraints, such as data storage capacity and download rates.

This simple example causes minimal impact on the individual spacecrafts.
And it offers services based upon availability of the resources without modifications in the space-to-ground links.
We have different satellites cooperating built up by using different technologies, from different organizations \cite{Golkar2015fss}.
New terrestrial DCPs can enter the network, increasing the demand and the regions of interest, i.e., global.
Ways to measure the quality of the service, as FSS, must be considered, even more with the effects of new arriving constituent systems.

\subsection{Setting up the requirements}

We can derive some requirements for the constellation from the definition of the goals and objectives of the FSS.

Functional and extra-functional requirements should be derived, but how to do it once we do not control the constituent systems?
One approach for that is the approach driven with the unifying concept of operation, ConOps, and its requirements \cite{wertz1999smad}.
The federation as a single System of Systems, SoS, has unique characteristics and requirements that the sum of its parts should meet. At the same time, every single participating satellite still keeps its original primary designated mission \cite{Akhtyamov2019retrofiting}.

Back to the BEDCS/GOLDS example, we can use the ConOps characteristics \cite{wertz1999smad} to define some requirements for the idealized FSS:

\begin{itemize}
    \item Data Availability
    \begin{itemize}
        \item The data will be processed onboard for the EDC payload satellites, and on ground for SCD and CBERS satellites.
        \item The data must be available as soon as the Mission Center validates the acquired data.
        \item All the data is centralized at the Mission Center, located in Brazil.
    \end{itemize}
    
    \item Communication Architecture
    \begin{itemize}
        \item Each satellite must define its particular downlink data rates in compliance with its ground stations.
        \item The access link between any satellite and the ground stations must be enough to download all the data from one complete orbital period.
        \item The access link between any satellite and one DCP must be enough to upload all the DCP data available.
        \item The revisit time (\textit{time between two consecutive overpasses on the same target}) for one DCP must not be more than 1 hour.
    \end{itemize}
    
    \item Tasks, Scheduling and Control
    \begin{itemize}
        \item The use and control of the hosted payload, EDC, must respect the BEDCS/GOLDS decisions and the satellite resources availability.
        \item All the federated ground stations must be capable of controlling the hosted payload, EDC.
    \end{itemize}
    
    \item Timeline
    \begin{itemize}
        \item The BEDCS/GOLDS must have the flexibility to receive new satellites with the hosted payload, EDC, to its federation.
        \item The BEDCS/GOLDS must have the flexibility to retire satellites from its federation to maintain the federation quality of service.
    \end{itemize}
    
    \item Fault Management
    \begin{itemize}
        \item In case of a fault on an FSS constituent system, the Mission Center must be able to perform a reconfiguration on the FSS resources.
        \item The time for reconfiguration must not exceed one operation planning period (\textit{operational activity when all the operational procedures are planned for a specific period of time, i.e., overpasses, flight plans, calibration, etc.}).
    \end{itemize}

\end{itemize}

As a System of Systems, an FSS evolves. Its behavior can be defined in terms of its systems independence and interoperability or, to be more specific, retrofitting, which is the capability for systems to interoperate on-demand to meet mission objectives as soon as a new satellite arrives/leaves the federation \cite{Madni2012sos}.

\subsection{Constraints}

Even knowing the overall objectives of an FSS, some items can remain fuzzy, again, due to the heterogeneity of the constituent systems.
How to dynamically integrate unknown resources? Will the system correctly provide the services?
Looking at the limitations of our system is, sometimes, more productive. The constraints of the FSS can best formulate the boundaries of its solution space. 

The requirements presented earlier can be refined into more detailed requirements and resource constraints.
Data availability requirements create constraints on each constituent satellite on data storage and processing.
Communication Architecture characteristics derive constraints on bandwidth and data rates to download DCP data.
Power consumption and federation engagement time on available resources constrain satellite control and tasks scheduling.
The deployment/retirement of new satellites requires a capability of reconfiguration and retrofitting on the FSS.
Fault management requirements also influence the FSS configuration.

Extensibility demands the introduction of quality measures into the set of requirements. The respective required minimal and offered values for new constituent satellites decide their integration. , e.g., at least $90\%$ of all DCP coverage, minimum of $10\%$ engagement time for non-dedicated satellites, 2 GB DCP data storage capacity, 2W peak power consumption, one day revisit time, minimum 10 minutes ground station access time per day, at least one dedicated ground station and communication channel \cite{vidal2021global}.

\begin{figure}[h!]
    \centering
    \includegraphics[width=0.45\textwidth]{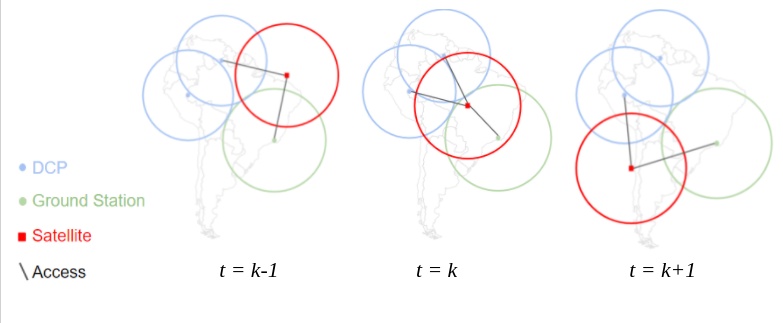}
    \caption{Coverage Constraint Problem for BEDCS.}
    \label{fig:problem}
\end{figure}

For example, Figure~\ref{fig:problem} shows the constraint of the BEDCS/GOLDS satellites coverage of DCP.

Note that, on BEDCS satellites, the access only exists if the satellite can 'view' the DCP and the Ground Station simultaneously, it is a COVERAGE constraint problem.

From now on, we will handle these constraints with the help of the mathematical paradigm called the Constraint Satisfaction Problem (CSP).
CSP is an approach that facilitates estimating a single solution, all solutions or the best solution for problems with limitations or conditions, defined into a domain set.

We can define each DCP as a region of interest (ROI) for the satellite and has its field of view (FOV).

\[ ROI = [ ROI_1 , ROI_2 , … , ROI_n ] \]

The same for the ground station(s):

\[ GrSt = [ GrSt_1 , GrSt_2 , … , GrSt_n ] \]

The Satellite has its FOV but it changes over time (orbit):

\[ Sat = [ fov(t) ] \]

If we deal with different satellites:

\[ Sat = [ fov_1(t) , fov_2(t) , … , fov_n(t)] \]

\begin{figure}[h!]
    \centering
    \includegraphics[width=0.45\textwidth]{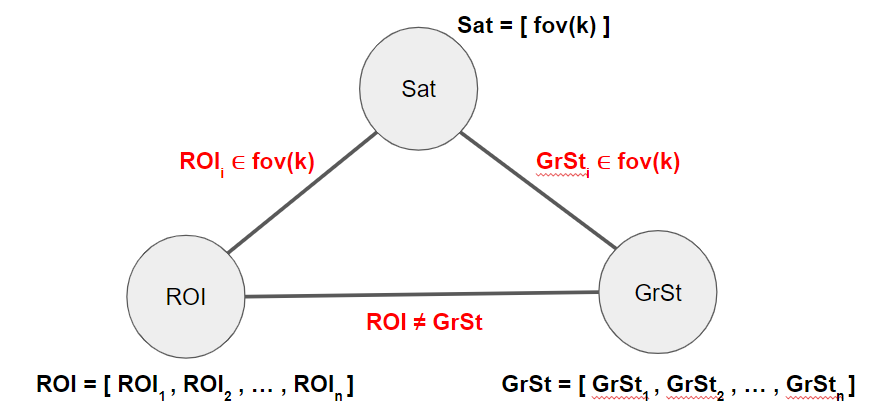}
    \caption{BEDCS Coverage Constraint Problem Model}
    \label{fig:model}
\end{figure}

Once the constraints are satisfied we have a successful access as can be viewed at the Figure~\ref{fig:model}.

As the satellite FOV changes over time, we will have an solution for each instant of time.

Given a time interval $t = [0, … , k]$ , the sum of these solutions for a specific ROI will give us the COVERAGE characteristic of the satellite for that ROI.

Moreover, if we have a constellation of $n$ satellites, we can derive the constellation COVERAGE for a specific ROI as the sum of each satellite set of solutions.

As we go to the GOLDS concept and the hosted EDC payload, the same problem of COVERAGE transforms itself into a CSP on data storage. The covered DCP networks upload to the satellite an amount of data associated with each ROI:

\[ Data = [ Data_1 , Data_2 , … , Data_n ] \]

Nevertheless, the satellite has two parameters, FOV and available storage at that specific time, $storage(t)$.

\[ Sat = [ ( fov(t), storage(t) ) ] \]

Again, if we deal with different satellites:

\[ Sat = [ ( fov_1(t), storage_1(t) ), ... , ( fov_n(t), storage_n(t) ) ] \]

\begin{figure}[!ht]
    \centering
    \includegraphics[width=0.45\textwidth]{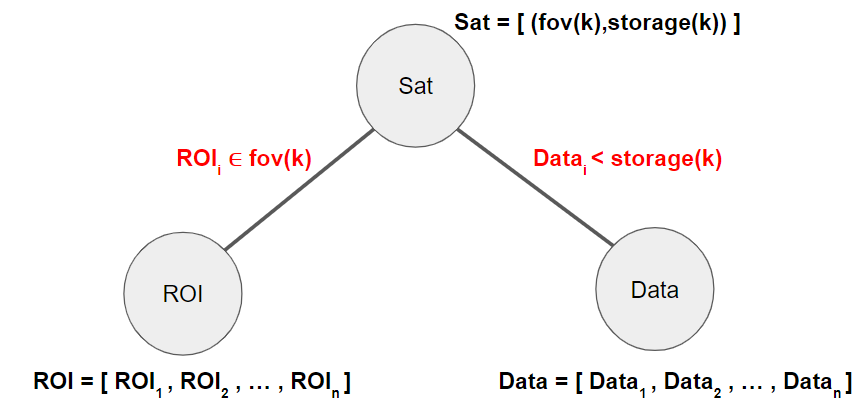}
    \caption{Data Access as Coverage Constraint Problem Model}
    \label{fig:modeldata}
\end{figure}

Once the constraints are satisfied, we have successful access, as can be viewed in Figure~\ref{fig:modeldata}.

As the satellite FOV changes over time, we will have a solution for each instant of time.

Given a time interval $t = [0, … , k]$ , the sum of these solutions for a specific ROI will give us the COVERAGE characteristic of the satellite for that ROI.

Moreover, if we have a constellation of $n$ satellites, we can derive the constellation COVERAGE for a specific ROI as the sum of each satellite set of solutions.

At the end, the sum of sets of solution for a specific ROI of the BEDCS and EDC satellites give us the COVERAGE of the GOLDS for that specific ROI.

So, evaluating the actual configuration characteristics (i.e. coverage, available satellites, ground stations, storage data, power, revisit time) and the desired configuration to achieve the expected quality of service is a must.
This reconfiguring capability configures an FSS emerging behavior by translating into a Constraint Optimization Problem instead of a satisfaction-only problem optimizing service provision with available resources.

\section{Conclusion}

Developing a Distributed Satellite System is a  challenge task.
Validating this idea using CubeSats can be game-changing for the next years.
Distributed CubeSat Systems have not yet been demonstrated in large scale, with exception of Planet and Spire over more than 40 proposed constellations and Federated CubeSats concepts have not yet flight.

We propose a possible protocol/process to validate the impacts on a heterogeneous federate CubeSat system of a new satellite or group of satellites deployed in orbit to work in this federation.
What is the problems inherent to this kind of system?
What to expect from the evolution of this SoS?
How much can be modeled once we do not have control over the constituent systems?

Using the BEDCS/GOLDS constellation as an example, we could translate some of the main characteristics of such constellation concepts.
We could also start to theorize over this complete satellite system (space, ground, and user segment) as an idea of cooperation and sharing of resources.

Another thing is how to simulate the system.
As a dynamic system, the federated CubeSat system and its constituent satellites time dependent, and the fulfillment of their constraints will also change over time.
Some tools can be used for that, orbit simulators are well-known but some work is necessary, e.g. \href{https://opensource.gsfc.nasa.gov/projects/GMAT/}{NASA General Mission Analysis Tool}.
Mainly we focus about the representation of the resources available at the system.

We still have significant work to do, not only on the modeling/model side but also on correctly picking the main attributes of this kind of constellation to better formulate the questions we have been asked during this whole paper.

\bibliographystyle{IEEEtran}
\bibliography{
    references/CubeSats.bib,
    references/CubeSatCapabilities.bib,
    references/references.bib,
    references/FSS-DSS-Constellations.bib
}

\end{document}